\documentclass[11pt]{article}

\usepackage[utf8]{inputenc}
\usepackage[T1]{fontenc}
\usepackage{lmodern}
\usepackage[margin=1in]{geometry}
\usepackage{amsmath,amssymb}
\usepackage{microtype}
\usepackage[numbers,sort&compress]{natbib}
\usepackage{hyperref}
\hypersetup{colorlinks=true,linkcolor=black,citecolor=black,urlcolor=blue}

\title{\texttt{comprisk}: A scikit-learn-compatible Python toolkit for
competing-risks survival analysis}

\author{%
  Sunny Yang\thanks{Corresponding author. University of Illinois
  Urbana-Champaign, USA. ORCID: 0009-0006-3160-0860.} \and
  Weiyan Zhao\thanks{Northumbria University, UK. ORCID: 0009-0005-7634-6715.}
  \and
  Wanqi Zhao\thanks{University of Calgary, Canada.}%
}

\date{July 2026}

\begin{document}
\maketitle

\begin{abstract}
Medical time-to-event data are frequently subject to \emph{competing
risks}, where the occurrence of one terminal event precludes the others and
standard survival methods that treat competing events as censoring yield
biased absolute-risk estimates. Valid analysis instead targets the
cause-specific cumulative incidence function (CIF). This methodology has been
available to applied researchers almost exclusively through R packages,
forcing Python-based machine-learning workflows into a Python-to-R round trip.
We present \texttt{comprisk}, a \texttt{scikit-learn}-compatible Python toolkit
that puts the canonical competing-risks methods behind one API: a scalable
competing-risks random survival forest, Fine--Gray subdistribution-hazard
regression and a penalized variant, cause-specific Cox regression, the
Aalen--Johansen CIF estimator, and Gray's $K$-sample test, together with
competing-risks-aware model evaluation. Every estimator is validated
numerically against its R reference implementation. The forest uses a histogram-based, \texttt{numba}-compiled
split kernel that fits 10--22$\times$ faster than \texttt{randomForestSRC} at
comparable discrimination on real clinical cohorts and scales
to $n = 10^{6}$ on a consumer CPU. \texttt{comprisk} is distributed on PyPI and
lets applied researchers run correct, scalable competing-risks analysis
without leaving the Python scientific stack.
\end{abstract}

\section{Summary}

Time-to-event data in medicine are frequently subject to \emph{competing
risks}: a patient may experience one of several mutually exclusive terminal
events (for example, death from heart failure versus death from other causes),
and the occurrence of one event precludes the others. Standard survival methods
that treat competing events as censoring produce biased absolute-risk estimates
\citep{austin2016, putter2007}. Correct analysis instead targets the
cause-specific cumulative incidence function (CIF) and models cause-specific or
subdistribution hazards.

\texttt{comprisk} is a Python package that brings the canonical toolkit of
competing-risks analysis into a single, \texttt{scikit-learn}-compatible
library. It provides a scalable competing-risks random survival forest together
with the standard regression and non-parametric estimators: Fine--Gray
subdistribution-hazard regression and a penalized variant,
cause-specific Cox regression, the Aalen--Johansen CIF estimator, and Gray's
$K$-sample test. It adds competing-risks-aware model evaluation: inverse
probability of censoring weighted (IPCW) time-dependent AUC and Brier score,
cause-specific concordance indices with closed-form confidence intervals, and
calibration curves. Every estimator is validated numerically against the
R reference implementations, so users obtain results consistent with
the published literature without leaving Python.

\section{Statement of need}

The canonical competing-risks toolkit lives in R: \texttt{cmprsk}
\citep{cmprsk} for Fine--Gray regression and Gray's test, \texttt{survival}
\citep{survival} for cause-specific Cox models, \texttt{crrp} \citep{crrp}
for penalized Fine--Gray (archived from CRAN in 2022), \texttt{riskRegression} \citep{riskRegression} for IPCW
scoring, and \texttt{randomForestSRC} \citep{ishwaran2008, ishwaran2014} for
competing-risks forests. Python has parts of it. \texttt{lifelines}
\citep{lifelines} provides an Aalen--Johansen cumulative-incidence estimator,
\texttt{scikit-survival} \citep{sksurv} has shipped a non-parametric
competing-risks CIF estimator since version 0.24.0, \texttt{pycox}
\citep{pycox} implements DeepHit \citep{deephit}, \texttt{statsmodels}
\citep{statsmodels} has a cause-specific cumulative incidence estimator, and
\texttt{hazardous} \citep{hazardous} adds a gradient-boosted cause-specific
incidence model together with IPCW Brier scores and a competing-risks
concordance index. What is missing is the classical layer, natively: no
competing-risks forest (the random survival forest in
\texttt{scikit-survival} is single-event), no Fine--Gray or other
subdistribution-hazard regression outside an \texttt{rpy2} wrapper around
\texttt{cmprsk}, no Gray's test, and no competing-risks time-dependent AUC or
closed-form inference for the concordance. An analyst who needs
those methods still leaves Python for R, which fragments reproducible
workflows and raises the barrier to correct competing-risks analysis.

\texttt{comprisk} supplies that modelling layer. Its central contribution is
a native-Python competing-risks random survival forest with competing-risks log-rank splitting
(both composite and cause-specific), Aalen--Johansen CIF and Nelson--Aalen
cumulative-hazard prediction, out-of-bag Breiman permutation variable
importance \citep{breiman2001}, Ishwaran minimal-depth variable selection, IPCW
cause-specific concordance \citep{wolbers2009, wolbers2014, uno2011}, and exact
cause-specific TreeSHAP attributions \citep{lundberg2020}. Time-dependent
Shapley explanations exist for single-event survival models
\citep{survshap}, but not for competing-risks cumulative incidence.

A histogram-based split kernel with \texttt{uint8}-binned features, just-in-time compiled with \texttt{numba}, gives
sub-linear wall-time growth in sample size: on real clinical cohorts it fits
10--22$\times$ faster than \texttt{randomForestSRC} at comparable discrimination
(both C-index $\approx 0.85$, each under its own native concordance scorer), and
on a synthetic feasibility benchmark it scales to $n = 10^{6}$ in roughly one
minute on a consumer CPU, where existing tools become memory-bound. Output is
bit-identical across thread counts for a fixed random seed. An optional
\texttt{equivalence="rfsrc"} mode reproduces \texttt{randomForestSRC}'s per-tree
mtry and nsplit random-number stream and exports the matched in-bag matrix, so
the two libraries can be compared tree by tree.

Alongside the forest, \texttt{comprisk} ships the regression and non-parametric
estimators an applied study needs, each validated to floating-point tolerance
against its R counterpart: Fine--Gray regression \citep{finegray1999} against
\texttt{cmprsk::crr}, penalized Fine--Gray (LASSO / ridge / elastic-net / MCP /
SCAD) against \texttt{crrp}, cause-specific Cox against \texttt{survival::coxph},
the Aalen--Johansen estimator \citep{aalen1978} against
\texttt{cmprsk::cuminc}, and Gray's test \citep{gray1988} against
\texttt{cmprsk::cuminc}. The evaluation module \texttt{score\_cr} /
\texttt{calibration\_cr} provides the IPCW time-dependent AUC, Brier score,
integrated variants, and calibration data that correspond to the
competing-risks mode of \texttt{riskRegression::Score}, and the
\texttt{concordance\_index\_ci} / \texttt{concordance\_index\_delta\_ci}
functions supply closed-form (bootstrap-free) confidence intervals and paired
model-comparison tests for the IPCW concordance based on its influence-function
variance \citep{wolbers2014}.

The intended audience is biostatisticians, epidemiologists, and data
scientists building risk-prediction models on competing-risks outcomes.

\section{Functionality}

The public API exposes:

\begin{itemize}
  \item \textbf{\texttt{CompetingRiskForest}}: competing-risks random
  survival forest with CIF / cumulative-hazard prediction, out-of-bag
  concordance scoring, permutation and minimal-depth variable importance, and
  exact TreeSHAP.
  \item \textbf{\texttt{FineGrayRegression}} and
  \textbf{\texttt{PenalizedFineGrayRegression}}: subdistribution-hazard
  regression with robust standard errors, and a cross-validated regularization
  path.
  \item \textbf{\texttt{CauseSpecificCox}}: cause-specific
  proportional-hazards regression.
  \item \textbf{\texttt{CumulativeIncidence}}: non-parametric
  Aalen--Johansen CIF estimation.
  \item \textbf{\texttt{gray\_test}}: Gray's $K$-sample test for equality of
  CIFs.
  \item \textbf{\texttt{score\_cr}}, \textbf{\texttt{calibration\_cr}},
  \textbf{\texttt{concordance\_index\_ci}}, and
  \textbf{\texttt{concordance\_index\_delta\_ci}}: competing-risks-aware
  model evaluation.
\end{itemize}

The package requires Python $\geq 3.10$ and depends only on the core
scientific-Python stack (\texttt{numpy}, \texttt{scipy}, \texttt{pandas},
\texttt{joblib}, \texttt{numba}, \texttt{scikit-learn}). It is distributed on
PyPI (\texttt{pip install comprisk}); the source code is developed at
\url{https://github.com/sunnyadn/comprisk} and archived on Zenodo
(DOI~\href{https://doi.org/10.5281/zenodo.19876282}{10.5281/zenodo.19876282}).
It is documented with a quickstart, worked notebooks, an autogenerated API
reference, and a benchmark dossier with reproduction scripts. Correctness is covered by a test
suite that includes property-based tests and paired-seed equivalence checks
against the R reference implementations.

\section{Implementation and design}

Continuous features are quantile-binned once into \texttt{uint8} codes,
and splits are searched over histograms of accumulated event counts rather than
by re-sorting observations at every node; this trades exact split thresholds for
histogram resolution (256 bins by default) and turns the per-node split scan
into a bounded, cache-friendly reduction. The split kernels are compiled with
\texttt{numba} and release the GIL, so trees are grown in parallel across cores
with \texttt{joblib} while the inner loops stay in native code. Fitted trees are
stored in a flat array layout with sparse leaf tables, which keeps the
serialized model compact and memory access local at prediction time.

Each estimator subclasses the \texttt{scikit-learn} \texttt{BaseEstimator}, so
they compose directly with \texttt{Pipeline}, \texttt{cross\_val\_score}, and
hyperparameter search without adapters. The forest exposes two independent
switches. \texttt{mode} selects the split search: the default histogram path
optimizes for speed and memory, and a \texttt{reference} path performs the
exact sort-based scan. \texttt{equivalence="rfsrc"} is a separate preset that
aligns the per-tree mtry and nsplit random-number stream with that of
\texttt{randomForestSRC} and exports the matched in-bag matrix, which lets a
user reproduce the comparison tree by tree. Both split paths are deterministic
and
bit-identical across thread counts for a fixed seed. An optional CUDA backend is
provided as a preview and falls back to the CPU path when a GPU is unavailable.

\section{Data availability}

The correctness test suite and a public synthetic two-cause Weibull benchmark
(shipped in the repository) can be reproduced by any user. The headline
real-cohort speed comparisons use a de-identified heart-failure electronic
health record cohort and the SEER breast-cancer registry; both are
access-restricted and require the respective data-use agreements, so those
specific tables are reproducible only by users with cohort access. Full
reproduction scripts and per-run provenance are provided in the repository's
benchmark dossier.

\section*{Generative AI disclosure}

AI-based coding assistants (Anthropic's Claude, accessed via the Claude Code
command-line interface) were used to assist with software development,
documentation, and preparation of this manuscript. All statistical methodology,
algorithmic and API design decisions, and the numerical validation of every
estimator against the established R reference implementations
(\texttt{randomForestSRC}, \texttt{cmprsk}, \texttt{crrp}, \texttt{survival},
\texttt{riskRegression}) were determined and carried out by the human authors,
who reviewed and edited all AI-assisted output and take full responsibility for
the correctness of the software and the accuracy of the claims made in this
paper. No AI tools were used for communication with editors or reviewers.

\section*{Acknowledgements}

We thank the maintainers of \texttt{randomForestSRC}, \texttt{cmprsk},
\texttt{crrp}, \texttt{survival}, and \texttt{riskRegression}, whose
implementations served as validation references during development.

\bibliographystyle{plainnat}
\bibliography{paper}

\end{document}